\documentclass[12pt]{article}
\newcommand{\be}{\begin{equation}}
\newcommand{\ee}{\end{equation}}
\newcommand{\bea}{\begin{eqnarray}}
\newcommand{\eea}{\end{eqnarray}}
\newcommand{\sptwo}{1.4}
\newcommand{\doublespace}{\edef\baselinestretch{\sptwo}\Large\normalsize}
\newcommand{\newsection}[1]{\section{#1}\setcounter{equation}{0}}

\newcounter{newapp}
\begin{document}
\vspace*{0.2in}
\begin{center}
{\large\bf Nonlinear realization of local symmetries of $AdS$ space}
\end{center}
\vspace{0.2in}
\begin{center}
{T.E. Clark}\footnote{e-mail address: clark@physics.purdue.edu}$~^a~~,~~${S.T. Love}\footnote{e-mail address: love@physics.purdue.edu}$~^{a,b}~~,~~${Muneto Nitta}\footnote{e-mail address: nitta@th.phys.titech.ac.jp}$~^c~~,~~${T. ter Veldhuis}\footnote{e-mail address: terveldhuis@macalester.edu}$~^d$\\
\end{center}
~\\
~\\
a. {\it Department of Physics, Purdue University, West Lafayette, IN 47907-1396}\\
~\\
b. {\it Fermi National Accelerator Laboratory, P.O. Box 500, Batavia, IL 60510}\\
~\\
c. {\it Department of Physics, Tokyo Institute of Technology, Tokyo 152-8551, Japan}\\
~\\
d. {\it Department of Physics \& Astronomy, Macalester College, Saint Paul, MN 55105-1899}
\vspace{0.2in}
~\\
~\\
~\\
\begin{center}
{\bf Abstract}
\end{center}
Coset methods are used to construct the action describing the dynamics associated with the spontaneous breaking of the local symmetries of $AdS_{d+1}$ space due to the embedding of an $AdS_{d}$ brane.  The resulting action is an $SO(2,d)$ invariant $AdS$ form of the Einstein-Hilbert action, which in addition to the $AdS_d$ gravitational vielbein, also includes a massive vector field localized on the brane. Its long wavelength dynamics is the same as a massive Abelian vector field coupled to gravity in $AdS_d$ space.
~\\
\pagebreak
\doublespace

\newsection{Introduction}

The dynamical degree of freedom of a probe $AdS_d$ brane embedded in an $AdS_{d+1}$ target space is the Nambu-Goldstone world volume field $\phi (x)$, with $x^\mu$ the $AdS_d$ world volume coordinates.  $\phi (x)$ describes the co-volume oscillatons of the brane in $AdS_{d+1}$ space.  Its long wavelength dynamics is given by the reparametrization invariant volume of the brane times the constant brane tension $\sigma$ and is encoded in an $AdS$ form \cite{Clark:2005ht} of the Nambu-Goto action as
\bea
\Gamma &=& -\sigma \int d^d x \det{e}=-\sigma \int d^d x \det{\bar{e}}\det{{N}} \cr
 &=&  -\sigma \int d^d x \det{\bar{e}} \cosh^d{\sqrt{m^2\phi^2}} \sqrt{1-\frac{{\cal D}_m \phi \eta^{mn}{\cal D}_n \phi}{\cosh^2{\sqrt{m^2\phi^2}}}}.
\label{nga}
\eea
The induced vielbein $e_\mu^{~m}$ has a product form, $e_\mu^{~m}=\bar{e}_\mu^{~n}N_n^{~m}$ where $\bar{e}_\mu^{~m}$ is the static background vielbein for the $AdS_d$ world volume of the brane which yields a background world volume Ricci tensor $\bar{R}_{\mu\nu}= m^2(d-1)\bar{g}_{\mu\nu}$ and hence a background Ricci scalar $\bar{R}=m^2 d(d-1)$ with $m^2 > 0$, while the $\phi$ dependent  $N_n^{~m}$ is given by
\bea
N_m^{~n}(x) &=& \delta_m^{~n} \cosh{(\sqrt{m^2\phi^2(x)})} \cr
 & & \quad +\left[ \sqrt{\left( \cosh^2{(\sqrt{m^2\phi^2(x)})} - {\cal D}_r \phi(x) \eta^{rs} {\cal D}_s \phi(x) \right)} \right. \cr
 & &\left. \qquad\qquad\qquad\qquad -\cosh{(\sqrt{m^2\phi^2(x)})} \right] \frac{{\cal D}_m \phi(x) {\cal D}^n \phi(x)}{({\cal D} \phi)^2(x)} ,
\eea
where the derivative ${\cal D}_m$ is defined as ${\cal D}_m = \bar{e}_m^{-1\mu} \partial_\mu$.
Expanding the action through terms bilinear in $\phi$ gives
\be
\Gamma =  -\sigma \int d^d x \det{\bar{e}}\left\{ 1+ \frac{1}{2}(m^2d)\phi^2 -\frac{1}{2}\partial_\mu \phi \bar{g}^{\mu\nu} \partial_\nu \phi +\cdots \right\}.
\ee
It is seen that the Nambu-Goldstone boson carries the $(E,s)=(d,0)$ representation \cite{Dusedau:1985ue} of $SO(2,d-1)$. That is, it has mass squared equal to $m^2 d$ and hence energy $d$ in units of $m^2$ while being spin zero \cite{Ivanov}-\cite{CNtV}.  It should be noted that the $m^2 =0$ case reproduces the massless bosonic brane Nambu-Goto action.

The brane action is invariant under a nonlinear realization of the $AdS_{d+1}$ target space global isometry group of transformations $SO(2,d)$.  In order to have invariance under general coordinate transformations, additional gravitational fields must be introduced.  The purpose of this paper is to construct the action of the world volume localized gravitational fields when the brane is embedded in curved space \cite{Karch:2000ct}-\cite{Salgado:2003rf}.  In short, the dynamics of the oscillating brane in curved space is described by a world volume localized massless graviton represented by a dynamical veilbein $e_\mu^{~m} (x)$ and a world volume localized vector field represented by a dynamical field $A_\mu (x)$. As a consequence of the Higgs mechanism \cite{Porrati:2001gx}-\cite{Porrati:2001db}, the vector field is massive.  The action for these fields is derived in a model independent manner using coset methods in which the $AdS_{d+1}$ local symmetry group $SO(2, d)$ is nonlinearly realized.  In section 2, the nonlinear local transformations of the Nambu-Goldstone fields are introduced via the coset method \cite{Coleman:sm}.  The locally covariant Maurer-Cartan one-form building blocks for the invariant action are obtained along with the  introduction of the dynamical veilbein and vector fields.  Derivatives of these Maurer-Cartan world volume vectors that are covariant with respect to local Lorentz and Einstein transformations are defined using the spin and related affine connections.  In section 3, these covariant derivatives are used to construct the low energy locally $SO(2, d)$ invariant action.  Exploiting the spontaneously broken local (pseudo-) translation and Lorentz transformations, the action is transformed to and analyzed in the unitary gauge.  The physical degrees of freedom so obtained are the dynamical world volume veilbein and massive vector field.
\pagebreak

\newsection{The Coset Construction}

The embedding of $AdS_{d}$ space spontaneously breaks the symmetry group of the $AdS_{d+1}$ space from $SO(2,d)$ to $SO(2,d-1)$.  The low energy action governing the dynamics of the Nambu-Goldstone modes associated with the symmetry breakdown can be constructed using coset methods.  This technique begins by introducing the coset element $\Omega \in SO(2,d)/SO(1,d-1)$ where $SO(1,d-1)$ corresponds to the Lorentz structure (stability) group of transformations in $AdS_d$.  The $AdS_d = SO(2,d-1)/SO(1,d-1)$ coordinates, $x^\mu$, act as parameters for pseudo-translations in the world volume and are part of the coset so that
\be
\Omega (x) = e^{ix^\mu  P_\mu} e^{i\phi(x)Z} e^{iv^\mu (x) K_\mu},
\label{coset}
\ee
The $SO(2,d)$ generators can be expressed in terms of the unbroken $SO(1,d-1)$ Lorentz subgroup representation content of the $SO(2,d-1)$ symmetry group of the brane.  The unbroken $SO(2,d-1)$ symmetry group is generated by the subgroup Lorentz transformation generators $M_{\mu\nu}$, where $\mu , \nu = 0,1,2,\ldots , d-1$ and the pseudo-translations in $AdS_d$ space with charges $P_\mu $.  The remaining charges are the generating elements of the $SO(2,d)/SO(2,d-1)$ coset.  They are the broken $SO(2,d)$ symmetry transformation charges.  $Z$ generates the broken $SO(2,d)$ pseudo-translations in the co-volume direction normal to the brane, while  $K_\mu$ generates the broken $AdS_{d+1}$ Lorentz transformations.  Thus the $SO(2,d)$ algebra can be written in terms of the $P_\mu$, $M_{\mu\nu}$, $Z$ and $K_\mu$ charges as \cite{Clark:2005ht}
\bea
\left[M_{\mu\nu} , M_{\rho\sigma} \right] &=& -i \left( \eta_{\mu\rho} M_{\nu\sigma} -\eta_{\mu\sigma} M_{\nu\rho} +\eta_{\nu\sigma} M_{\mu\rho} -\eta_{\nu\rho} M_{\mu\sigma} \right) \cr
\left[M_{\mu\nu} , P_{\lambda} \right] &=& i \left( P_\mu \eta_{\nu\lambda} - P_\nu \eta_{\mu\lambda} \right) \cr
\left[M_{\mu\nu} , K_{\lambda} \right] &=& i \left( K_\mu \eta_{\nu\lambda} - K_\nu \eta_{\mu\lambda} \right) \cr
\left[M_{\mu\nu} , Z \right] &=& 0 \cr
\left[P_\mu , P_\nu \right] &=& -i m^2 M_{\mu\nu} \cr
\left[K_\mu , K_\nu \right] &=& i M_{\mu\nu} \cr
\left[P_\mu , K_\nu \right] &=& i \eta_{\mu\nu} Z \cr
\left[P_\mu , Z \right] &=& -i m^2 K_\mu \cr
\left[Z , K_\mu \right] &=& i P_\mu .
\label{Algebra}
\eea
The coset so defined corresponds to a particular choice of coordinates, specifically denoted as $x^\mu$, for the $AdS_d$ world volume.  The fields are also defined as functions of $x^\mu$.  The Nambu-Goldstone field $\phi (x)$ along with $v^\mu (x)$ act as the remaining coordinates needed to parametrize the coset manifold $SO(2,d)/SO(2,d-1)$. 

Left multiplication of the coset elements $\Omega$ by an $SO(2,d)$ group element $g$ which is specified by local infinitesimal parameters $\epsilon^\mu (x), z(x), b^\mu(x) , \lambda^{\mu\nu}(x)$ so that 
\be
g(x) = e^{i\epsilon^\mu(x) P_\mu} e^{iz(x)Z} e^{ib^\mu(x) K_\mu} e^{\frac{i}{2} \lambda^{\mu\nu}(x) M_{\mu\nu}},
\ee
results in transformations of the space-time coordinates and the Nambu-Goldstone fields according to the general form \cite{Coleman:sm}
\be
g(x)\Omega(x) = \Omega^\prime(x^\prime) h(x) .
\label{leftmult}
\ee
The transformed coset element, $\Omega^\prime$,  is a function of the transformed world volume coordinates and the total variations of the fields so that
\be
\Omega^\prime = e^{ix^{\prime\mu}  P_\mu} e^{i\phi^\prime(x^\prime)Z} e^{iv^{\prime\mu} (x^\prime) K_\mu},
\ee 
while $h$ is a field dependent element of the stability group $SO(1,d-1)$:
\be
h= e^{\frac{i}{2} \theta^{\mu\nu}(x) M_{\mu\nu}} .
\label{hele}
\ee
Exploiting the algebra of the $SO(2,d)$ charges displayed in equation (\ref{Algebra}), along with extensive use of the Baker-Campbell-Hausdorf formulae, the local $SO(2, d)$ transformations are obtained as
\bea
x^{\prime\mu}  &=&  \left[ 1 -z(x)\sqrt{m^2} \tanh{\sqrt{m^2\phi^2}}\frac{\sin{\sqrt{4m^4  x^2}}}{\sqrt{m^2 x^2}}\right]x^\mu  -\lambda^{\mu\nu}(x) x_\nu \cr
 & &\qquad + \left[ P_{L}^{\mu\nu}(x) + \sqrt{m^2 x^2}\cot{\sqrt{m^2 x^2}}P_{T}^{\mu\nu}(x)\right] \epsilon_\nu (x) \cr
 & &\qquad\qquad + \frac{\tanh{\sqrt{m^2\phi^2}}}{\sqrt{m^2}} \left[ \cos{\sqrt{m^2 x^2}}P_{L}^{\mu\nu}(x) + \frac{\sqrt{m^2 x^2}}{\sin{\sqrt{m^2 x^2}}}P_{T}^{\mu\nu}(x)\right] b_\nu (x)\cr
 & & \cr
\phi^\prime (x^\prime) &=& \phi (x) + z(x)\cos{\sqrt{m^2 x^2}} + b_\mu (x) x^\mu \frac{\sin{\sqrt{m^2 x^2}}}{\sqrt{m^2 x^2}}\cr
 & & \cr
v^{\prime \mu} (x^\prime) &=& v^\mu (x) -\lambda^{\mu\nu}(x) v_\nu -\frac{m^2}{2}\frac{\tan{\sqrt{m^2 x^2/4}}}{\sqrt{m^2 x^2/4}}(\epsilon^\mu (x) x^\nu -\epsilon^\nu (x) x^\mu)v_\nu \cr
 & &\qquad -z(x)\frac{m^2}{\cosh{\sqrt{m^2\phi^2}}} \frac{\sin{\sqrt{m^2 x^2}}}{\sqrt{m^2 x^2}}\left[ P_{L}^{\mu\nu}(v) + \sqrt{v^2}\coth{\sqrt{v^2}}P_{T}^{\mu\nu}(v)\right] x_\nu \cr
 & &\qquad\qquad + \sqrt{m^2} \frac{\tan{\sqrt{m^2 x^2/4}}}{\sqrt{m^2 x^2}}\tanh{\sqrt{m^2 \phi^2}} (b^\mu (x) x^\nu - b^\nu (x) x^\mu ) v_\nu \cr
 & & +\frac{1}{\cosh{\sqrt{m^2\phi^2}}} \left[ P_{L}^{\mu\nu}(v) + \sqrt{v^2}\coth{\sqrt{v^2}}P_{T}^{\mu\nu}(v)\right] \cr
 & & \qquad\qquad\qquad\qquad\qquad\qquad\times\left[ \cos{\sqrt{m^2x^2}} P_{L\nu\rho}(x) + P_{T\nu\rho}(x)\right] b^\rho (x) \cr
 & & \cr
\theta^{\mu\nu}(x) &=& \lambda^{\mu\nu}(x) + \frac{m^2}{2} \frac{\tan{\sqrt{m^2x^2/4}}}{\sqrt{m^2x^2/4}}(\epsilon^\mu (x) x^\nu - \epsilon^\nu (x) x^\mu ) \cr
 & &\qquad -z(x)\frac{m^2}{\cosh{\sqrt{m^2\phi^2}}} \frac{\sin{\sqrt{m^2 x^2}}}{\sqrt{m^2 x^2}}(v^\mu x^\nu -v^\nu x^\mu ) \frac{\tanh{\sqrt{v^2/2}}}{\sqrt{v^2}}\cr
 & &\qquad\qquad -\sqrt{m^2} \frac{\tan{\sqrt{m^2 x^2/4}}}{\sqrt{m^2 x^2}}\tanh{\sqrt{m^2 \phi^2}} (b^\mu (x) x^\nu - b^\nu (x) x^\mu )\cr
 & & -\frac{1}{\cosh{\sqrt{m^2\phi^2}}} \frac{\tanh{\sqrt{v^2/2}}}{\sqrt{v^2}}  \cr
 & &\qquad\qquad\qquad \times\left[ \cos{\sqrt{m^2x^2}} P_{L}^{\mu\rho}(x)   b_\rho (x) v^\nu + P_{T}^{\mu\rho}(x) b_\rho (x) v^\nu - (\mu \leftrightarrow \nu ) \right].\cr
 & & 
\label{variations}
\eea
Here the transverse and longitudinal projectors for $x^\mu$ are defined as
\bea
P_{T\mu\nu}(x)&=& \eta_{\mu\nu} -\frac{x_\mu x_\nu}{x^2} \cr
P_{L\mu\nu}(x)&=&\frac{x_\mu x_\nu}{x^2} 
\eea
and $\eta_{\mu\nu}$ is the metric tensor for $d$--dimensional Minkowski space having signature $(+1,-1,\ldots,-1)$.  In the above, the indices are raised, lowered and contracted using $\eta_{\mu\nu}$.
Both Nambu-Goldstone fields $\phi$ and $v^\mu$ transform inhomogeneously under the broken local translations $Z$ and broken local Lorentz transformations $K_\mu$.  Thus these broken transformations can be used to transform to the unitary gauge in which both $\phi$ and $v^\mu$ vanish.  This will be done in section 3 in order to exhibit the physical degrees of freedom in a more transparent fashion.

The nonlinearly realized $SO(2,d)$ transformations induce a coordinate and field dependent general coordinate transformation of the world volume space-time coordinates.  From the $x^\mu$coordinate transformation given above, the $AdS_{d+1}$ general coordinate Einstein transformation for the world volume space-time coordinate differentials is given by
\be
dx^{\prime \mu} = dx^\nu {G}_\nu^{~\mu} (x),
\label{dxprime}
\ee
with ${G}_\nu^{~\mu}(x) = \partial x^{\prime \mu}/\partial x^\nu$.  The $SO(2,d)$ invariant interval can be formed using the metric tensor ${g}_{\mu\nu}(x)$ so that $ds^2 = dx^\mu {g}_{\mu\nu}(x) dx^\nu = ds^{\prime 2} = dx^{\prime \mu} {g}^\prime_{\mu\nu}(x^\prime) dx^{\prime \nu}$ where the metric tensor transforms as 
\be
{g}^\prime_{\mu\nu} (x^\prime) = {G}_\mu^{-1\rho}(x) {g}_{\rho\sigma}(x) {G}_\nu^{-1\sigma}(x) .
\label{gprime}
\ee

The form of the vielbein (and hence the metric tensor) as well as the locally $SO(2,d)$
covariant derivatives of the Nambu-Goldstone boson fields and the spin connection can be extracted from the locally covariant Maurer-Cartan one-form, $\Omega^{-1}D\Omega$, which can be expanded in terms of the generators as 
\bea
\Omega^{-1} D\Omega &\equiv& \Omega^{-1} (d +i \hat{E})\Omega \cr
 &=& i\left[ \omega^m P_m + \omega_Z Z +\omega^m_K K_m +\frac{1}{2}\omega_M^{mn} M_{mn}\right].
\eea
Here Latin indices $m,n = 0,1,\ldots,d-1$, are used to distinguish tangent space local Lorentz transformation properties from world volume Einstein transformation properties which are denoted using Greek indices. In what follows Latin indices are raised and lowered with use of the Minkowski metric tensors, $\eta^{mn}$ and $\eta_{mn}$, while Greek indices are raised and lowered with use of the curved $AdS_d$ metric tensors, $g^{\mu\nu}$ and $g_{\mu\nu}$.  Since the Nambu-Goldstone fields vanish in the unitary gauge it is useful to exhibit the one-form gravitational fields in terms of their pseudo-translated form
\be
\hat{E} = e^{+ix^\mu P_\mu} E e^{-ix^\mu P_\mu} .
\ee
The world volume one-form gravitational fields $E$ have the expansion in terms of the charges as
\be
E= E^m P_m +A Z + B^m K_m +\frac{1}{2}\gamma^{mn} M_{mn} .
\ee
Similarly expanding $\hat{E}$ as 
\be
\hat{E}= \hat{E}^m P_m +\hat{A} Z + \hat{B}^m K_m +\frac{1}{2}\hat{\gamma}^{mn} M_{mn} ,
\ee
one finds the various fields are related according to 
\bea
\hat{E} &=& \left[\cos{\sqrt{m^2 x^2}}P^m_{Tn}(x) +P^m_{Ln}(x)\right] E^n -\frac{\sin{\sqrt{m^2 x^2}}}{\sqrt{m^2 x^2}} \gamma^{mn}x_n \cr
\hat{A} &=& A \cos{\sqrt{m^2 x^2}} - \frac{\sin{\sqrt{m^2 x^2}}}{\sqrt{m^2 x^2}} B^m x_m \cr
\hat{B^m}&=& \left[P^m_{Tn}(x) +\cos{\sqrt{m^2 x^2}} P^m_{Ln}(x)\right] B^n +m^2 \frac{\sin{\sqrt{m^2 x^2}}}{\sqrt{m^2 x^2}} A x^m \cr
\hat{\gamma}^{mn} &=& \gamma^{mn} - m^2 \frac{\sin{\sqrt{m^2 x^2}}}{\sqrt{m^2 x^2}} \left( E^m x^n - E^n x^m \right) \cr
 & & -\left( \cos{\sqrt{m^2 x^2}} -1\right) \left[ \gamma^{ms} P^n_{Ls} (x) -\gamma^{ns} P^m_{Ls}(x) \right]. 
\eea

Defining the one-form gravitational fields to transform as a gauge field so that
\be
\hat{E}^\prime (x^\prime) = g(x)\hat{E}(x) g^{-1}(x) -ig(x)dg^{-1}(x),
\ee
the covariant Maurer-Cartan one-form transforms analogously to the way it varied for global transformations:
\be
\omega^\prime(x^\prime) = h(x)\omega (x) h^{-1}(x) +h(x)dh^{-1}(x),
\ee
with $h= e^{\frac{i}{2}\theta^{mn}(x) M_{mn}}$.  Expanding in terms of the $SO(2,d)$ charges, the individual one-forms transform according to their local Lorentz nature
\bea
\omega^{\prime m}(x^\prime) &=&  \omega^n (x)\Lambda^{~m}_{n}(\theta (x)) \cr
\omega_Z^\prime (x^\prime)&=& \omega_Z (x)\cr
\omega^{\prime m}_K (x^\prime)&=&  \omega^n_K (x)\Lambda^{~m}_{n}(\theta (x))\cr
\omega^{\prime mn}_M (x^\prime)&=& \omega^{rs}_M (x)\Lambda^{~m}_{r}(\theta (x))\Lambda^{~n}_{s}(\theta (x))-d\theta^{mn}(x) .
\label{oneformvari}
\eea
For infinitesimal transformations, the local Lorentz transformations are $\Lambda^{~m}_{n}(\theta (x)) = \delta^{~m}_{n} + \theta^{~m}_{n}(x)$, while the infinitesimal local $SO(2,d)$ transformations of the gravitational one-forms take the form
\bea
\hat{E}^{\prime m} &=& \hat{E}^m + \hat{\gamma}^{mn}\epsilon_n - z\hat{B}^m +b^m \hat{A} -\lambda^{mn}\hat{E}_n -d\epsilon^m \cr
\hat{A}^\prime &=& \hat{A} -\epsilon_m \hat{B}^m + b_m \hat{E}^m - dz \cr
\hat{B}^{\prime m} &=& \hat{B}^m + \epsilon^m m^2 \hat{A} -z m^2 \hat{E}^m + \hat\gamma^{mn} b_n -\lambda^{mn}\hat{B}_n -db^m   \cr
\hat\gamma^{\prime mn} &=& \hat\gamma^{mn} + m^2( \epsilon^m \hat{E}^n - \epsilon^n \hat{E}^m) - (b^m \hat{B}^n -b^n \hat{B}^m) \cr
 & & \qquad\qquad + ( \lambda^{mr}\hat\gamma^n_{~r} - \lambda^{nr} \hat\gamma^m_{~r}) -d\lambda^{mn}.
\eea

Using the Feynman formula for the variation of an exponential operator in conjunction with the Baker-Campell-Hausdorff formulae, the individual world volume one-forms appearing in the above decomposition of the covariant Maurer-Cartan one-form are secured as 
\bea
\omega^m &=& dx^\mu e_\mu^{~m}\cr
  &=& dx^\mu{\cal E}_\mu^{~n} N_n^{~m} \cr
\omega_Z &=& dx^\mu \omega_{Z\mu} \cr
 &=& dx^\mu \cosh{\sqrt{v^2}} {\cal E}_\mu^{~m} \left[ - v_m \frac{\tanh{\sqrt{v^2}}}{\sqrt{v^2}}\cosh{\sqrt{m^2\phi^2}} + {\cal E}_m^{-1\nu} \left( \partial_\nu\phi + A_\nu \right) \right]\cr
\omega_K^m &=& dx^\mu \omega_{K\mu}^m \cr
 &=& \left[P_L^{mn} (v) + \frac{\sinh{\sqrt{v^2}}}{\sqrt{v^2}} P_T^{mn} (v) \right] \left( dv_n -(\bar\omega_{Mnr} + \gamma_{nr})v^r \right) \cr
 & & +\left[ P_{L}^{mn}(v) + \cosh{\sqrt{v^2}}P_{T}^{mn}(v) \right] \left[(\bar\omega_n +E_n ) \sqrt{m^2}\sinh{\sqrt{m^2\phi^2}}\right. \cr
 & &\left. \qquad\qquad\qquad\qquad\qquad\qquad\qquad\qquad\qquad\qquad\qquad + B_n \cosh{\sqrt{m^2\phi^2}} \right]\cr
\omega_M^{mn} &=& dx^\mu \omega_{M\mu}^{mn}\cr
 &=& (\bar\omega_M^{mn} +\gamma^{mn} ) - \cosh{\sqrt{m^2 \phi^2}} \frac{\sinh{\sqrt{v^2}}}{\sqrt{v^2}} \left[ B^m v^n -B^n v^m \right] \cr
 & &+\sqrt{m^2} \frac{\sinh{\sqrt{v^2}}}{\sqrt{v^2}} \sinh{\sqrt{m^2 \phi^2}} \left[ v^m P_{Ts}^n (v) -v^n P_{Ts}^m (v) \right] ( \bar\omega^s + E^s ) \cr
 & &+\left[ \cosh{\sqrt{v^2}} -1 \right] \left[\frac{v^m dv^n - v^n dv^m}{v^2} \right. \cr
 & &\left. \qquad\qquad\qquad\qquad\qquad -\left( P_{Lr}^m (v) (\bar\omega_M^{nr} + \gamma^{nr} ) -P_{Lr}^n (v) (\bar\omega_M^{mr} + \gamma^{mr} ) \right)\right] .\cr
 & & 
\label{MCOne-form}
\eea
In these expressions, the background $AdS_d$ covariant coordinate differential, $\bar\omega^m$, and spin connection, $\bar\omega_M^{mn}$, are obtained from the $AdS_d$ coordinate one-form $(e^{-ix^m P_m})d (e^{ix^n P_n})=i[\bar\omega^m P_m + \frac{1}{2}\bar\omega_M^{mn} M_{mn}]$ as
\bea
\bar\omega^m &=& \frac{\sin{\sqrt{m^2x^2}}}{\sqrt{m^2x^2}} P_{T}^{mn}(x) dx_n  +P_{L}^{mn}(x) dx_n \cr
\bar\omega_M^{mn} &=& \left[ \cos{\sqrt{m^2x^2}} -1\right] \frac{(x^m dx^n -x^n dx^m)}{x^2}.
\label{AdSdoneform}
\eea
The background differential $\bar\omega^m$ is related to the $x^\mu$ world volume coordinate differential by the background veilbein $\bar{e}_\mu^{~m}(x)$ as
\be
\bar\omega^m = dx^\mu \bar{e}_\mu^{~m}(x) .
\ee  
Using equation (\ref{AdSdoneform}) along with $d=dx^\mu \partial_\mu^x$, the background veilbein is obtained as
\be
\bar{e}_\mu^{~m}(x) = \frac{\sin{\sqrt{m^2x^2}}}{\sqrt{m^2x^2}} P_{T\mu}^{~~~m}(x)  +P_{L\mu}^{~~~m}(x) .
\label{ebar}
\ee
The covariant coordinate differential $\omega^m$ is related to the world volume coordinate differential $dx^\mu$ by the vielbein $e_\mu^{~m}$ which in turn can be written in a factorized form as the product of the dynamic vielbein ${\cal E}_\mu^{~m}$ 
and the Nambu-Goto vielbein $N_n^{~m}$
\bea
{\cal E}_\mu^{~m} &=& \bar{e}_\mu^{~m} + E_\mu^{~m} + B_\mu^{~m} \frac{\tanh{\sqrt{m^2 \phi^2}}}{\sqrt{m^2}}  \cr
N_n^{~m} &=& \cosh{\sqrt{m^2 \phi^2}} \left[ P_{Tn}^{~~m} (v) + \cosh{\sqrt{v^2}} P_{Ln}^{~~m} (v) \right]  \cr
 & & \qquad\qquad - \cosh{\sqrt{v^2}} {\cal E}_n^{-1 \nu} (\partial_\nu \phi + A_\nu ) v^m \frac{\tanh{\sqrt{v^2}}}{\sqrt{v^2}} .
\label{vng}
\eea

The one-forms and their covariant derivatives are the building blocks of the locally $SO(2,d)$ invariant action.  Indeed a $m^{\rm th}$-rank contravariant local Lorentz and $n^{\rm th}$-rank covariant Einstein tensor, $T^{m_1\cdots m_m}_{\mu_1\cdots \mu_n}$ is defined to transform as \cite{Utiyama:1956sy}
\be
T^{\prime m_1^\prime\cdots m_m^\prime}_{\mu_1^\prime\cdots \mu_n^\prime}(x^\prime) = G_{\mu_1^\prime}^{-1\mu_1}(x)\cdots G_{\mu_n^\prime}^{-1\mu_n}(x) T^{m_1\cdots m_m}_{\mu_1\cdots \mu_n}(x)
\Lambda^{~m_1^\prime}_{m_1} (\theta (x))\cdots \Lambda^{~m_m^\prime}_{m_m} (\theta (x)). 
\ee
For example, the veilbein transforms as $e_\mu^{\prime m} (x^\prime) = G_{\mu}^{-1\nu}(x)
e_\nu^{~n}(x)\Lambda^{~m}_{n} (\theta (x))$.  Hence, the veilbein and its inverse can be used to convert local Lorentz indices into world volume indices and vice versa.  Since the Minkowski metric, $\eta_{mn}$, is invariant under local Lorentz transformations the metric tensor $g_{\mu\nu}$
\be
g_{\mu\nu} = e_\mu^{~m} \eta_{mn} e_\nu^{~n} ,
\ee
is a rank 2 Einstein tensor.  It can be used to define covariant Einstein tensors given contravariant ones.  Likewise, the Minkowski metric can be used to define covariant local Lorentz tensors given contravariant ones.  

Since the Jacobian of the $x^\mu \rightarrow x^{\prime \mu}$ transformation is simply
\bea
d^dx^\prime &=& d^d x ~\det{{G}},  
\eea
it follows that $d^dx^\prime ~\det{e^\prime} (x^\prime) = d^dx ~\det{e} (x)$ since $\det{\Lambda} = 1$.
Thus an $SO(2,d)$ invariant action is constructed as
\be
\Gamma = \int d^d x \det{e(x)} {\cal L}(x),
\ee
with the Lagrangian an invariant ${\cal L}^\prime (x^\prime) = {\cal L}(x)$.  The invariants that make up the Lagrangian can be found by contracting the indices of tensors with the appropriate vielbein, its inverse and the Minkowski metric.  For example $\omega_{Z\mu} g^{\mu\nu} \omega_{Z\nu}$ is an invariant term used to construct the action.

Besides products of the covariant Maurer-Cartan one-forms, their covariant derivatives can also be used to construct invariant terms of the Lagrangian.  The covariant derivative of a general tensor can be defined using the affine and related spin connections.  Consider the covariant derivative of the Lorentz tensor $T^{mn}$
\be
\nabla_\rho T^{mn} = \partial_\rho T^{mn} -\omega_{M\rho r}^{m} T^{rn}-\omega_{M\rho r}^{n} T^{mr} .
\ee
Since the spin connection transforms inhomogeneously according to equation (\ref{oneformvari}), the covariant derivative of $T^{mn}$ transforms homogeneously again 
\be
(\nabla_\rho T^{mn})^\prime =G_\rho^{-1\sigma} (\nabla_\sigma T^{rs})\Lambda_r^{~m}\Lambda_s^{~n} .
\ee
Converting the Lorentz index $n$ to a world index $\nu$ using the vielbein, the covariant derivative for mixed tensors is obtained
\bea
\nabla_\rho T^{m\nu} &\equiv & e_n^{-1\nu} \nabla_\rho T^{mn} = \partial_\rho T^{m\nu} -\omega_{M\rho}^{mr}T_r^{~\nu} + \Gamma_{\sigma\rho}^\nu T^{m\sigma} ,
\eea
where the spin connection $\omega_{M\rho}^{mn}$ and $\Gamma_{\sigma\rho}^\nu$ are related according to \cite{Utiyama:1956sy}
\be
\Gamma_{\sigma\rho}^\nu = e_n^{-1\nu} \partial_\rho e_\sigma^{~n} -e_n^{-1\nu} \omega_{M\rho}^{nr} e_\sigma^{~s} \eta_{rs}.
\ee
(Note that this relation as well follows from the requirement that the covariant derivative of the vielbein vanishes, $\nabla_\rho e_\mu^{~m} =0$.)
Applying the above to the Minkowski metric Lorentz 2-tensor yields the formula relating the affine connection $\Gamma^\rho_{\mu\nu}$ to derivatives of the metric
\bea
\nabla_\rho \eta^{mn} &=& \partial_\rho \eta^{mn} -\omega_{M\rho r}^{m} \eta^{rn}-\omega_{M\rho r}^{n} \eta^{mr} = -\omega_{M\rho}^{mn} -\omega_{M\rho}^{nm}\cr
 &=& 0 \cr
 &=& e_\mu^{~m}e_\nu^{~n} \nabla_\rho g^{\mu\nu} \cr
 &=& e_\mu^{~m}e_\nu^{~n}\left( \partial_\rho g^{\mu\nu} +\Gamma_{\sigma\rho}^\mu g^{\sigma\nu} + \Gamma_{\sigma\rho}^\nu g^{\mu\sigma} \right).
\eea
The solution to this equation yields the affine connection in terms of the derivative of the metric \cite{Utiyama:1956sy} (the space is torsionless, hence the connection is symmetric $\Gamma^\rho_{\mu\nu} = \Gamma^\rho_{\nu\mu}$)
\be
\Gamma^\rho_{\mu\nu} = \frac{1}{2}g^{\rho\sigma}\left[ \partial_\mu g_{\sigma\nu} +\partial_\nu g_{\mu\sigma} - \partial_\sigma g_{\mu\nu}\right].
\ee

Finally a covariant field strength two-form can be constructed out of the inhomogeneously transforming spin connection $\omega_{M\mu}^{mn}$
\bea
F^{mn} &=& d\omega_M^{mn} +\eta_{rs} \omega_M^{mr}\wedge \omega_M^{ns} .
\eea
Expanding the forms yields the field strength tensor
\be
F^{mn}_{\mu\nu} = \partial_\mu \omega_{M\nu}^{mn} -\partial_\nu \omega_{M\mu}^{mn}+\eta_{rs} \omega_{M\mu}^{mr} \omega_{M\nu}^{ns} -\eta_{rs} \omega_{M\nu}^{mr} \omega_{M\mu}^{ns} .
\ee
It can be shown that $F^{mn}_{\mu\nu}=e^{-1n\sigma}e_\rho^{~m}R^\rho_{~\sigma\mu\nu}$ where $R^\rho_{~\sigma\mu\nu}$ is the Riemann curvature tensor
\be
R^\rho_{~\sigma\mu\nu} = \partial_\nu \Gamma^\rho_{\sigma\mu} -\partial_\mu \Gamma^\rho_{\sigma\nu} +\Gamma^\lambda_{\sigma\mu}\Gamma^\rho_{\lambda\nu} -\Gamma^\lambda_{\sigma\nu}\Gamma^\rho_{\lambda\mu} .
\ee
The Ricci tensor is given by $R_{\mu\nu} = R^\rho_{\mu\nu\rho}$ and hence the scalar curvature is an invariant
\be
R = g^{\mu\nu} R_{\mu\nu} = - e^{-1\mu}_m e_n^{-1\nu} F^{mn}_{\mu\nu} .
\ee
\pagebreak

\newsection{The Invariant Action}

The covariant derivatives of the Maurer-Cartan one-forms provide additional building blocks out of which the invariant action is to be constructed.  For example the covariant derivatives of $\omega_{Z\nu}$ and $\omega_{K\nu}^n$ yield the mixed tensors
\bea
\nabla_\mu \omega_{Z\nu} &=& \partial_\mu \omega_{Z\nu} - \Gamma_{\mu\nu}^\rho \omega_{Z\rho} \cr
\nabla_\mu \omega_{K\nu}^n &=& \partial_\mu \omega_{K\nu}^n -\Gamma_{\mu\nu}^\rho \omega_{K\rho}^n -\omega_{M\mu}^{nr} \omega_{K\nu}^s \eta_{rs} .
\eea
So proceeding, the invariant action describing the curved $AdS_d$ brane embedded in curved $AdS_{d+1}$ space has the general low energy form
\bea
\Gamma &=& \int d^d x \det{e} \left\{ \Lambda + \kappa^2 R +\frac{1}{2} \omega_{Z\mu} \left[(M^2 + \xi R)g^{\mu\nu} + \zeta R^{\mu\nu}\right]  \omega_{Z\nu} \right.\cr
 & &\left.  \right. \cr
 & &\left.- \frac{1}{2}\nabla_\mu \omega_{Z\nu} \nabla_\rho \omega_{Z\sigma}\left[ Z_1( g^{\mu\rho}g^{\nu\sigma} -  g^{\mu\sigma}g^{\nu\rho}) +Z_2 g^{\mu\nu}g^{\rho\sigma}\right]\right.\cr
 & &\left.  \right. \cr
 & &\left.+\frac{1}{2} \omega_{K\mu}^m \omega_{K\nu}^n \left[ a e_m^{-1\mu} e_n^{-1\nu} +b e_n^{-1\mu} e_m^{-1\nu} +c g^{\mu\nu} \eta_{mn} \right] \right. \cr
 & &\left.  \right. \cr
 & &\left. -\omega_{K\mu}^m \nabla_\nu \omega_{Z\rho} \left[ \alpha e_m^{-1\mu} g^{\nu\rho} + \beta e_m^{-1\nu} g^{\mu\rho} + \gamma e_m^{-1\rho} g^{\mu\nu} \right] \right\} .
\eea
Many invariant terms are possible. The above includes a reduced set of terms which leads to a consistent effective theory.  The model can be further simplified by setting the parameters $\xi$ and $\zeta$ to zero. On the other hand, due to the Higgs mechanism, the parameter $M$ cannot be zero and is an independent scale in the theory.  Since the massive vector $A_\mu$ is a Proca field, it can be consistently quantized by further setting $Z_2$ to zero. Moreover, exploiting the identity $\partial_\mu (\det{e}~ T^\mu )= \det{e}~ \nabla_\mu T^\mu$ along with the chain rule for covariant differentiation, integration by parts has been used to eliminate redundant terms.  A term of the form $\det{e} e_m^{-1\mu} \omega_{K\mu}^m$ has also been excluded from the action since, when $\omega_{K\mu}^m$ is eliminated as below, it will not result in any new terms.
 
The action is independent of any terms containing derivatives acting on $\omega_{K\mu}^m$.  Hence varying the action with respect to $\omega_{K\mu}^m$ yields an algebraic identity relating it to $\nabla_\mu \omega_{Z\nu}$ as
\bea
 & &\omega_{K\mu}^m \left[ a e_m^{-1\mu} e_\sigma^{~r} +b \delta_\sigma^\mu \delta_m^r + c e^{-1\mu r} e_{\sigma m}\right] \cr
 &=& \nabla_\nu \omega_{Z\rho} \left[ \alpha g^{\nu\rho} e_\sigma^{~r} + \beta e^{-1\rho r} \delta_\sigma^\nu + \gamma e^{-1\nu r} \delta_\sigma^\rho \right] .
\eea
Introducing $\omega_{K\mu}^m = e^{-1\nu m} B_{\mu\nu}$ allows the solution
\bea
B_{\sigma\rho} &=& g_{\rho\sigma} \frac{1}{(b+c)}\left[ \alpha - a \frac{(\alpha d +\beta + \gamma)}{(ad +b +c)}\right] g^{\mu\nu} \nabla_\mu \omega_{Z\nu} \cr
 & &+\frac{(\beta +\gamma)}{2(b+c)}\left[ \nabla_\sigma \omega_{Z\rho}+ \nabla_\rho \omega_{Z\sigma} \right] 
+ \frac{(\beta -\gamma)}{2(b-c)}\left[ \nabla_\sigma \omega_{Z\rho}-\nabla_\rho \omega_{Z\sigma} \right] .
\eea
Substituting this back into the action allows $\omega_{K\mu}^m$ to be eliminated in favor of terms involving the vielbein and $\omega_{Z\mu}$ yielding
\bea
\Gamma &=& \int d^d x \det{e} \left\{ \Lambda + \kappa^2 R +\frac{1}{2}\omega_{Z\mu} \left[(M^2 +\xi R) g^{\mu\nu} + \zeta R^{\mu\nu}\right] \omega_{Z\nu} \right.\cr
 & &\left.  \right. \cr
 & &\left.\qquad - \frac{1}{2}Z_1\nabla_\mu \omega_{Z\nu} \nabla_\rho \omega_{Z\sigma}( g^{\mu\rho}g^{\nu\sigma} -  g^{\mu\sigma}g^{\nu\rho}) \right\}
,
\eea
where the form of the contractions of the product of two $\nabla_\mu \omega_{Z\nu}$ terms are similar to those of the initial action and hence the constants have just been redefined and the effective $Z_2$ has been set to zero.  Exploiting the form of the covariant derivative of the $Z$ one-form in order to define the anti-symmetric field strength tensor $F_{\mu\nu}$, 
\be
F_{\mu\nu}=(\nabla_\mu \omega_{Z\nu} -\nabla_\nu \omega_{Z\mu}) = (\partial_\mu \omega_{Z\nu} -\partial_\nu \omega_{Z\mu}) ,
\ee
the action becomes
\bea
\Gamma &=& \int d^d x \det{e} \left\{ \Lambda + \kappa^2 R +\frac{1}{2}\omega_{Z\mu} \left[(M^2 +\xi R) g^{\mu\nu} + \zeta R^{\mu\nu}\right] \omega_{Z\nu} \right.\cr
 & &\left.  \right. \cr
 & &\left.\qquad\qquad\qquad - \frac{Z_1}{4}F_{\mu\nu}g^{\mu\rho}g^{\nu\sigma} F_{\rho\sigma}  \right\}.
\label{effaction}
\eea

According to equation (\ref{variations}), $\phi$ and $v^m$ transform inhomogeneously under the broken translation and Lorentz transformation local transformations.  Hence we now fix the unitary gauge defined by  $\phi =0= v^m$. So doing, the covariant one-forms take a simplified form
\bea
\omega^m &=& dx^\mu e_\mu^{~m} =dx^\mu {\cal E}_\mu^{~m} = dx^\mu (\bar{e}_\mu^{~m} + E_\mu^{~m}) \cr
\omega_Z &=& dx^\mu  A_\mu \cr
\omega_K^m &=& dx^\mu B_\mu^{~m}\cr
\omega_M^{mn} &=& (\bar\omega_M^{mn} +\gamma^{mn} ) =dx^\mu (\bar\omega_{M\mu}^{mn} +\gamma_\mu^{mn} ).
\label{MCOne-formUnitary}
\eea
Note that, in this gauge, equation (\ref{vng}) reduces to ${\cal E}_\mu^{~m} =\bar{e}_\mu^{~m} + E_\mu^{~m}$ and $N_b^{~a} = \delta_b^{~a}$.  Consequently the vielbein $e_\mu^{~m} = {\cal E}_\mu^{~b} N_b^{~a} = \bar{e}_\mu^{~m} + E_\mu^{~m}$ and thus depends only on the gravitational fluctuation field, $E_\mu^{~m}$, about the $AdS_d$ background vielbein $\bar{e}_\mu^{~m}$ and is independent of the vector field. As such, the $\det{e} $ gives no contribution to the vector mass even though it is the source of Nambu-Goldstone boson kinetic term in the model with spontaneously broken global isometry. Instead, the mass of the vector, $M$, is a completely new scale arising from an  independent monomial. This realization of the Higgs mechanism is strikingly different from what occurs when gauging internal symmetries. In that case, when the symmetry is made local, the Nambu-Goldstone boson kinetic term gets replaced by the square of the covariant derivative containing the vector connection. In unitary gauge, the Nambu-Goldstone field vanishes leaving the residual vector mass term whose scale is set by the Nambu-Goldstone decay constant, a scale already present in the global model.  

The action, equation (\ref{effaction}), reduces to that of a massive vector field coupled to a gravitational field with cosmological constant
\be
\Gamma = \int d^d x \det{e} \left\{ \Lambda + \kappa^2 R - \frac{Z_1}{4}F_{\mu\nu}g^{\mu\rho}g^{\nu\sigma} F_{\rho\sigma}+\frac{1}{2}A_\mu \left[(M^2 +\xi R) g^{\mu\nu} + \zeta R^{\mu\nu}\right] A_\nu\right\},
\ee
with the field strength tensor $F_{\mu\nu}$ for the vector field
\be
F_{\mu\nu} = \partial_\mu A_\nu - \partial_\nu A_\mu .
\ee
The action was constructed by considering gravitational fluctuations about a background static $AdS$ space and describes the embedding of a brane in that curved space.  The world volume action of the brane is equivalent to that of a world volume gravitational field Einstein-Hilbert action, with corresponding cosmological constant as dictated by the field equations evaluated on the $AdS$ background, and the action for a massive vector field in that gravitating space.  Furthermore, the action can equally well be used to describe a bosonic brane embedded in a space gravitating about a background Minkowski space by taking the limit $m^2 \rightarrow 0$.  The vector field remains massive with the mass $M$ still being an independent scale.  Setting the parameters $\xi$ and $\zeta$ to zero, the world volume action for a brane embedded in curved space has the form of a massive Abelian gauge theory coupled to gravity.
~\\
~\\
\noindent The work of TEC and STL was supported in part by the U.S. Department of Energy under grant DE-FG02-91ER40681 (Task B) while MN was supported by the Japan Society for the Promotion of Science under the Post-Doctoral Research Program. STL thanks the hospitality of the Fermilab theory group during his sabbatical leave while this project was undertaken.  TtV would like to thank the theoretical physics groups at Purdue University and the Tokyo Institute of Technology for their hospitality during visits while this work was being completed.

\end{document}